\documentclass[final,5p,times,twocolumn]{elsarticle}
\usepackage{import}

\usepackage{amssymb} 
\usepackage{lineno} 
\usepackage{booktabs}
\usepackage{threeparttable} 
\usepackage{footnote}

\usepackage[labelfont=bf,justification=raggedright,singlelinecheck=false]{caption}
\usepackage{subfigure} 
\usepackage{multirow} 
\usepackage[dvipsnames]{xcolor} 
\usepackage{ulem} 

\usepackage{array}

\journal{Nuclear Physics A}

\begin{document}

\begin{frontmatter}
    \title{Trigger slow control system of the Belle II experiment}
    \author[1]{C.-H.~Kim\corref{cor1}}
\author[1]{Y.~Unno}
\author[1]{B.~G.~Cheon}

\author[1,2]{S.H.~Kim}
\author[1,2]{I.S.~Lee}

\author[3]{T.~Koga}
\author[3]{Y.-T.~Lai}
\author[3]{Y.~Iwasaki}
\author[3]{S.~Yamada}
\author[3,4]{M.~Nakao}

\author[5]{H.~Nakazawa}

\author[6]{E.-J.~Jang}
\author[6]{S.-K.~Choi}

\author[7]{T.~Konno}

\author[8,3]{D.~Liventsev}

\author[9]{S.-H.~Park}
\author[9]{Y.-J.~Kwon}

\author[10]{O.~Hartbrich}

\author[11]{M.~Ritzert}

\cortext[cor1]{Corresponding author, hun4341@hanyang.ac.kr}

\address[1]{Department of Physics and Institute of Natural Sciences, Hanyang University, Seoul 04763}
\address[2]{Institute for Basic Science, Daejeon 34126}
\address[3]{High Energy Accelerator Research Organization (KEK), Tsukuba 305-0801}
\address[4]{The Graduate University for Advanced Studies (SOKENDAI), Hayama 240-0193}
\address[5]{Department of Physics, National Taiwan University, Taipei 10617}
\address[6]{Gyeongsang National University, Jinju 52828}
\address[7]{Kitasato University, Sagamihara 252-0373}
\address[8]{Virginia Polytechnic Institute and State University, Blacksburg, Virginia 24061}
\address[9]{Yonsei University, Seoul 03722}
\address[10]{University of Hawaii, Honolulu, Hawaii 96822}
\address[11]{University of Heidelberg, 68131 Mannheim}
    \begin{abstract}
\begin{linenumbers}
The Belle II experiment at the SuperKEKB $e^{+}e^{-}$ collider in KEK, Japan, started physics data-taking with a complete detector from early 2019 with the primary physics goal of probing new physics in heavy quark and lepton decays. An online trigger system is indispensable for the Belle II experiment to reduce the beam background events associated with high electron and positron beam currents without sacrificing the target physics-oriented events. During the Belle II operation upon beam collision, the trigger system must be consistently controlled and its status must be carefully monitored in the process of data acquisition against unexpected situations. For this purpose, we have developed a slow control system for the Belle II trigger system. Around seventy thousand configuration parameters are saved in the Belle II central database server for every run when a run starts and stops. These parameters play an essential role in offline validation of the quality of runs. Around three thousand real-time variables are stored in the Belle II main archiving server, and the trend of some of these variables are regularly used for online and offline monitoring purposes. Various operator interface tools have been prepared and used. When the configuration parameters are not correctly applied, or some of the processes are unexpectedly terminated, the slow control system detects it, stops the data-taking process, and generates an alarm. In this article, we report how we constructed the Belle II trigger slow control system, and how we successfully managed to operate during its initial stage.
\end{linenumbers}
\end{abstract}
\end{frontmatter}

\section{Introduction}

The Belle II experiment~\citep{TA10} at the SuperKEKB~\citep{KA18} $e^{+}e^{-}$ collider in KEK, Japan, started physics data-taking with a complete detector from early 2019 with the primary physics goal of probing new physics in heavy quark and lepton decays. Belle II and SuperKEKB are the successors of Belle~\citep{AA02} and KEKB~\citep{SK03}, which have discovered Charge-Parity (CP) violation in B meson decay in 2001 to confirm the Cabibbo-Kobayashi-Maskawa (CKM) mechanism and led to the Nobel Prize by M. Kobayashi and T. Maskawa in 2008.

The target instantaneous luminosity of SuperKEKB is 40 times higher, and the target integrated luminosity is 50 times higher than its predecessor (Table~\ref{tab:lum}). The Belle II detector is designed to cope with the increase of the event rate as well as the harsh beam background at the target luminosity. Likewise, the trigger system has been upgraded to have robustness and flexibility, and a corresponding slow control system has been prepared to ensure physics data-taking.

In this paper, we describe the trigger slow control system of the Belle II experiment. Section 2 briefly describes the Belle II detector and its trigger and slow control systems. Section 3 describes how we successfully establish the framework for the trigger slow control system. In section 4, we describe the operator interface tools based on the trigger slow control system framework. In section 5, we explain how we utilize the trigger slow control system during the operation, and we summarize in section 6.

\begin{table}[!htb]
\centering
\caption{The KEKB/Belle achieved and the SuperKEKB/Belle II target luminosities.}
\label{tab:lum}
\begin{tabular*}{8.0cm} {@{\extracolsep{\fill} } lrr}
\toprule
\cmidrule(r){1-2}
& \multicolumn{1}{c}{Instantaneous}                         & \multicolumn{1}{c}{Integrated}                        \\
& \multicolumn{1}{c}{($10^{34} \ \mathrm{cm^{-2} s^{-1}}$)} & \multicolumn{1}{c}{($\mathrm{ab^{-1}}$)} \\
\midrule
KEKB/Belle          &   2.1 &   1.0 \\
SuperKEKB/Belle II  &    80 &    50 \\
\bottomrule
\end{tabular*}
\end{table}

\section{Belle II experiment}

\subsection{Belle II detector}
The Belle II detector consists, from the Interaction Point (IP) outward, of an inner silicon tracker comprising a PiXel Detector (PXD), a Silicon Vertex Detector (SVD), a Central Drift Chamber (CDC), two dedicated particle identification systems, the Aerogel Ring-Imaging CHerenkov detector (ARICH) in the forward endcap and the Time-Of-Propagation (TOP) detector in the barrel region, an Electromagnetic CaLorimeter (ECL), a superconducting solenoid for a homogeneous magnetic field of $1.5 \ \mathrm{T}$, and a $K_L$ and a Muon detector (KLM). A detailed description of the Belle II detector is in reference~\citep{TA10}.

Each of the Belle II sub-detector systems has its front-end readout system, which, except PXD, sends data to the unified back-end readout system. The back-end system consists of readout boards called COmmon Pipelined Platform for Electronics Readout (COPPER)~\citep{TH05} and readout servers~\citep{SY17}. The COPPERs receive sub-detector front-end data, including trigger data through high-speed optical links.

\subsection{Belle II trigger system}
The Belle II Level 1 (L1) online trigger (TRG) is required to achieve almost 100 $\mathrm{\%}$ trigger efficiency for $\Upsilon(4S)\rightarrow B\bar{B}$ events and nearly high efficiency for other physics processes of interest within 30 $\mathrm{kHz}$ maximum trigger rate under the harsh beam background environment of SuperKEKB. The readout system requires the trigger latency to be under 4.4 $\mathrm{\mu s}$. The cross section and trigger rate of physical processes at the Belle II target luminosities are listed in Table~\ref{tab:trg_target}.

\begin{table}[!htb]
\centering
\begin{threeparttable}
\caption{The total cross section from physics processes at the $\Upsilon(4S)$ energy region and expected trigger rates at the peak luminosity of $80 \times \ 10^{34} \ \mathrm{cm^{-2} s^{-1}}$~\citep{SHK17, YI11}.}
\label{tab:trg_target}
\begin{tabular*}{8.5cm} {@{\extracolsep{\fill} } lrr}
\toprule
Physics process                                          & Cross section ($\mathrm{nb}$) & Rate ($\mathrm{Hz}$)                       \\
\midrule
$e^{+}e^{-}\rightarrow \Upsilon(4S)\rightarrow B\bar{B}$ &                               1.1 &                      880\tnote{ } \\
$e^{+}e^{-}\rightarrow q\bar{q}$                         &                               3.4 &                     2700\tnote{ } \\
$e^{+}e^{-}\rightarrow \mu^{+}\mu^{-}$                   &                               1.1 &                      880\tnote{ } \\
$e^{+}e^{-}\rightarrow \tau^{+}\tau^{-}$                 &                               0.9 &                      720\tnote{ } \\
Bhabha\tnote{a}                                          &                              44.0 &                      350\tnote{c} \\
$\gamma\gamma$\tnote{a}                                  &                               2.4 &                       19\tnote{c} \\
$e^{+}e^{-}\rightarrow e^{+}e^{-} + 2 \gamma$ \tnote{a}\tnote{b}                      &                              13.0 &                    10000\tnote{d} \\
\midrule
Total                                                    &                               1.1 &              $\sim$15000\tnote{ } \\
\bottomrule
\end{tabular*}
\begin{tablenotes}
\item [a] $\theta_{lab} \geq 17^{\mathrm{\circ}}$
\item [b] $p_t \geq 0.1 \ \mathrm{GeV}$
\item [c] Pre-scaled by factor of 1/100.
\item [d] Estimated from the Belle level 1 trigger rate.
\end{tablenotes}
\end{threeparttable}
\end{table}

\begin{figure}[htb]
\centering
\includegraphics[width=0.8\columnwidth]{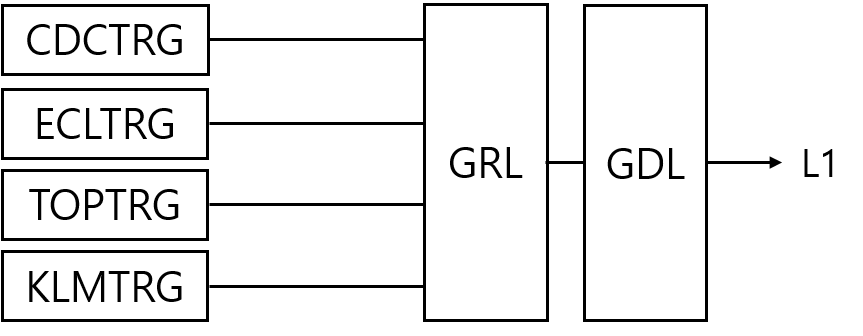}
\caption{A schematic overview of the Belle II trigger system~\citep{YI11}. The trigger system consists of the CDC / ECL / TOP and KLM triggers. The GRL collects all signals from each sub-trigger and delivers its matching information output to the GDL. The GDL generates a final decision with the matching information and sub-trigger outputs. L1 is the final level 1 trigger decision.}
\label{fig:SYS} 
\end{figure}

A schematic overview of the Belle II hardware trigger system is shown in Fig.~\ref{fig:SYS}. There are four sub-trigger systems, which are the CDC trigger, the ECL trigger, the TOP trigger, and the KLM trigger. The CDC trigger provides two- and three- dimensional (2D and 3D) charged track information~\citep{JG16}. The ECL trigger provides total energy, cluster information, and Bhabha identification information of electromagnetic particles~\citep{SHK17, BGC10}. The TOP trigger provides precise timing and hit topology information. The KLM trigger provides muon track information~\citep{YI11}.


The Global Reconstruction Logic (GRL) generates matching information based on all the sub-trigger outputs. The matching information and sub-trigger outputs are delivered to the Global Decision Logic (GDL), and the GDL makes the final trigger decision, called the L1 decision. Finally, the L1 decision is forwarded to the Belle II Data AcQuisition (DAQ) system~\citep{SY15}. For stable operation of the trigger system upon beam collision, a reliable controlling and monitoring system is mandatory.

\subsection{Belle II slow control system}
\label{subsec:slc}

The primary tasks of the Belle II slow control systems are to manage operations of sub-systems, to provide and store configuration parameters to sub-systems, and to collect status and environmental conditions of sub-systems. In order to accomplish these tasks, three fundamental sub-components have been developed; run control, log collector, and operator interface~\citep{TK15}. Four core techniques are used to build the components; the network communication based on EPICS~\citep{EPICS} and NSM2~\citep{MN00}, databases based on PostgreSQL~\citep{SQL}, the Graphical User Interface (GUI) based on the Control System Studio (CSS)~\citep{CSS} and Process Variable (PV)~\citep{SHP18} archiving via EPICS archiver appliances~\citep{EPICSAPPL}. The Belle II DAQ / slow control software package, named ``daq\_slc," contains all components. Each component is briefly explained below.

First, network communication between slow control processes is handled via EPICS and a software package named ``Network Shared Memory 2 (NSM2)", an advanced version of NSM~\citep{MN00}, which was used at the Belle experiment. A daemon process of the NSM2 takes care of all communication over the network segments~\citep{TK15}. Based on NSM2, we prepared NSM2 nodes, which are building blocks of the network communication between sub-systems through the daemon process. An NSM2 node reads some values from a hardware system, converts it to NSM2 format, and distributes it to the Belle II network. The NSM2 makes it possible for each operator to control systems remotely with minimized interference with each other.

Second, the PostgreSQL package is used for managing configuration parameters and logging information~\citep{TK15}.

Third, GUI tools are based on the Control System Studio (CSS), which is an Eclipse-based tool for creating control systems~\citep{CSS}. The CSS is originally designed for EPICS variables, but it also accepts NSM2 variables through a custom plugin on the CSS and an interface process, named nsm2socket~\citep{TK15}.

\begin{figure}[hp]
\centering
\includegraphics[width=0.85\columnwidth]{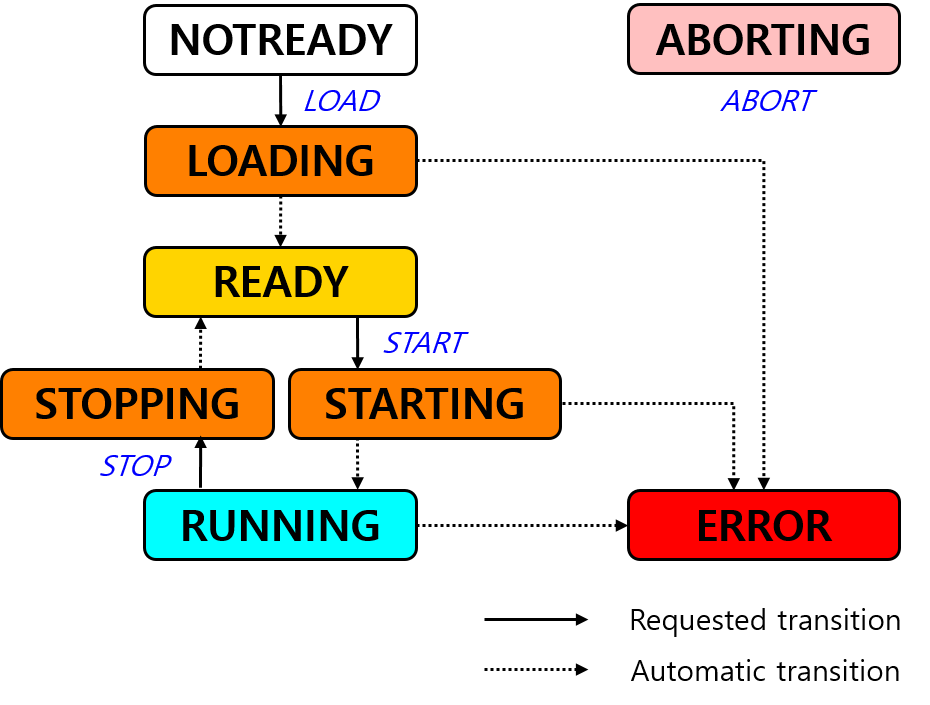}
\caption{Run control state diagram~\citep{TK15}. This diagram shows how the Belle II experiment handles a run. The round boxes denote the status of an NSM2 node, and the texts outside of the boxes are requests. A node status can be checked and changed by operators. The ABORTING state is reached from almost any states by an ABORT request.}
\label{fig:runcon} 
\end{figure}

\begin{table}[!htb]
\centering
\caption{List of run control requests.}
\label{tab:req}
\begin{tabular}{p{20mm} p{60mm}}
\toprule
Request & Meaning                                  \\      
\midrule
LOAD     & Load parameters                         \\
START    & Start a run                              \\
STOP     & Stop a run                               \\
ABORT    & Abort the state transition and reset the state \\
\bottomrule
\end{tabular}
\end{table}

\begin{table}[!htb]
\centering
\caption{List of run control states.}
\label{tab:rcstate}
\begin{tabular}{p{20mm} p{60mm}}
\toprule
State   & Meaning                             \\      
\midrule
NOTREADY & Parameters are not loaded yet      \\
LOADING  & Parameters are being loaded        \\
READY    & Ready to start a run               \\
STOPPING & Run is being stopped             \\
STARTING & Run is being started             \\
RUNNING  & Run is ongoing                   \\
ERROR    & Problem detected                   \\
ABORTING & Abort sequence is ongoing          \\
\bottomrule
\end{tabular}
\end{table}

The last one is the EPICS archiver appliance, which is an implementation of an archiver for EPICS control system~\citep{EPICSAPPL}. The EPICS archiver appliance records the history of PVs in real-time. In order to record the history of NSM2 variables, a process called ``nsm2cad" converts NSM2 variables to EPICS PVs because the EPICS Archiver Appliance can record EPICS PVs only~\citep{SHP18}.

The run control system manages the DAQ systems and sub-detector systems via NSM2 to follow the state diagram shown in Fig.~\ref{fig:runcon}~\citep{TK15}. The run control requests (requested by operators) and the run control states are listed in Table~\ref{tab:req} and~\ref{tab:rcstate}, respectively.
\begin{figure*}[t]
\centering
\includegraphics[width=0.9\linewidth]{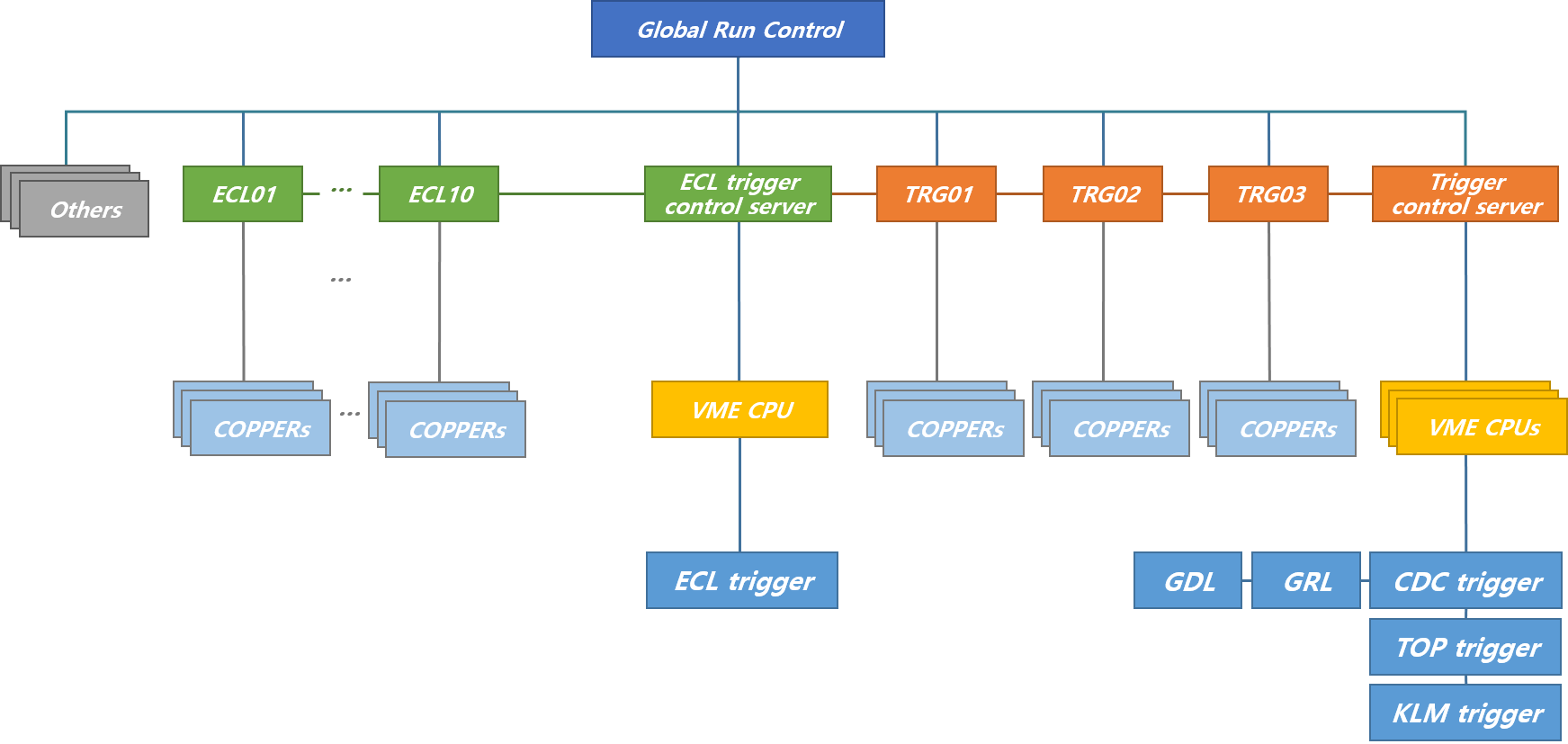}
\caption{The trigger slow control system server structure. This diagram shows how the trigger slow control related servers are connected to the Belle II network. ``Others" in the gray-colored box denote servers related to other sub-detectors; PXD, SVD, CDC, ARICH, TOP, and KLM. The TRG01/02/03 and the ECL01 to 10 servers are COPPER control severs. All the sub-trigger except the ECL trigger is connected to the trigger control server via corresponding VME CPUs. The ECL trigger system is connected to the ECL trigger control server via a VME CPU.}
\label{fig:SRV}
\end{figure*}

\begin{figure}[htb]
\centering
\includegraphics[width=0.95\columnwidth]{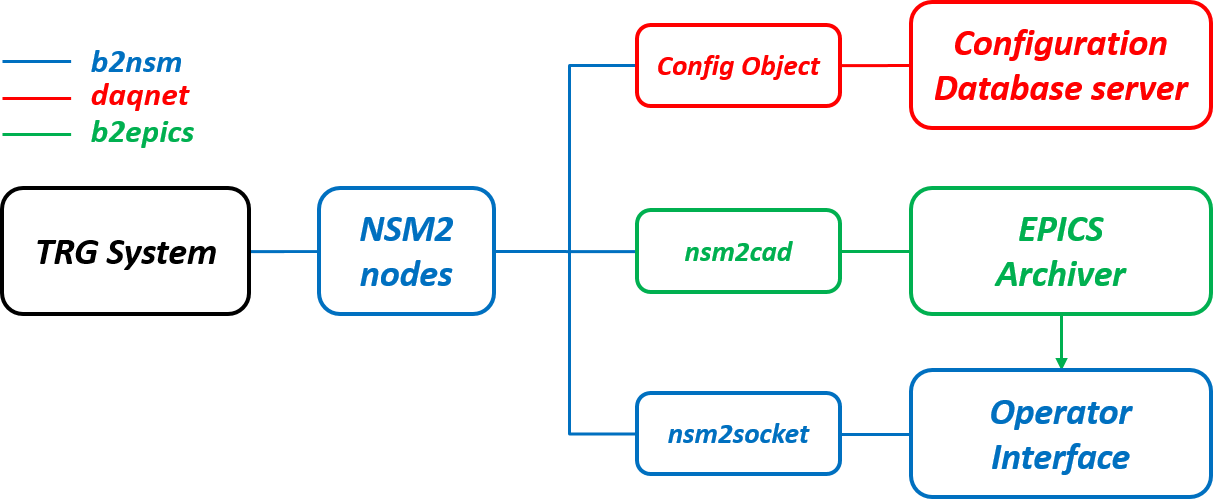}
\caption{NSM2 nodes are connected to three network lines: b2nsm, daqnet, and b2epics. The b2nsm is for NSM2 communication, the daqnet is for connections to the database server, and b2epics is for EPICS PVs. ``Config object," which is a C++ object of an NSM2 node, converts NSM2 variables to suitable for the configuration database format. ``nsm2cad" and ``nsm2socket" are explained in the previous section.}
\label{fig:3types}
\end{figure}

\section{Trigger slow control system framework}
\label{sec:ovv}

The stable operation of the trigger system is essential for the steady data acquisition of the Belle II experiment. The trigger slow control system makes it possible to operate the trigger system efficiently and stably. In order to cooperate with the rest of the slow control system, the trigger slow control system is constructed upon the Belle II common daq\_slc framework, described in subsection~\ref{subsec:slc}. The primary tasks of the trigger slow control system are shown below.

\begin{enumerate}[(a)]
    \item Makes logs of problematic events and sends it to the central logging server.
    \item Saves experimental parameters automatically at every run start and stop.
    \item Provides real-time monitoring plots.
    \item Takes local run data.
    \item Troubleshooting.
\end{enumerate}

In order to accomplish these tasks, the trigger system is connected to the Belle II main run control, central database, and archiver servers. Figure~\ref{fig:SRV} shows the connections between the trigger system servers, and to the main run control server.

To communicate through NSM2, we run several NSM2 nodes on the servers of the trigger system. The NSM2 nodes collect values from sub-trigger systems and convert them to NSM2 variables. NSM2 variables are then converted into three formats: configuration database, archiver, and operator interface format (Fig.~\ref{fig:3types}). The purpose of the configuration database format is to put experimental parameters from hardware to the configuration database server and vice versa. The goal of the archiver format is to save the trend of the values and view their history later. The operator interface format aims to display values of interest on an operator panel or change hardware settings from an operator panel. We will explain each part further in the following subsections.

\subsection{Configuration database}
A run-recording process automatically stores 67,662 experimental parameters to the central database server by run number when a run starts/ends. We can also save frequently used experimental parameters to the central database server, and NSM2 nodes can retrieve them. The parameters are mainly firmware versions and setting parameters. Some slow control settings, such as NSM2 port settings and including/excluding NSM2 nodes in run control (RC) scheme information, are also stored. We use the stored parameters to validate the run quality. A log-collecting process gathers the status of the NSM2 node and sends them to the central database server. We can check when problems happened and chase the origin of them with the collected logs.

\subsection{Archiver}
We archive 3,303\footnote{The Belle II archiver server is capable of storing 80,000 values. The trigger group have 5,000 slots.} EPICS PVs to the Belle II main archiver server\footnote{The default archiving period is 10 seconds, and the minimum period is 1 second.}. These are mainly variables to judge the stability of trigger systems such as the temperatures of modules, the pedestal levels, the luminosity, and the trigger rates. Some of them are plotted on CSS and monitored by trigger experts during a run in real-time.

\subsection{Operator interface tools}
User-friendly, intuitive, and reliable GUI is highly required for stable operation. We use CSS, which is widely used in the high energy physics community. NSM2 variables can be read/written on CSS by nsm2socket process and CSS NSM2 plugin. They convert the NSM2 variables to be handled by CSS. We show a few examples of frequently used operator interface panels in the next sections in detail.

\begin{itemize}
    \setlength \itemsep{-0.2em}
    \item The trigger main control panel
    \item The Level 1 trigger plot
    \item The trigger local run control panel
    \item The GDL main panel
    \item The ECL trigger panels
\end{itemize}
\begin{figure*}[t]
\centering
\includegraphics[width=0.8\linewidth]{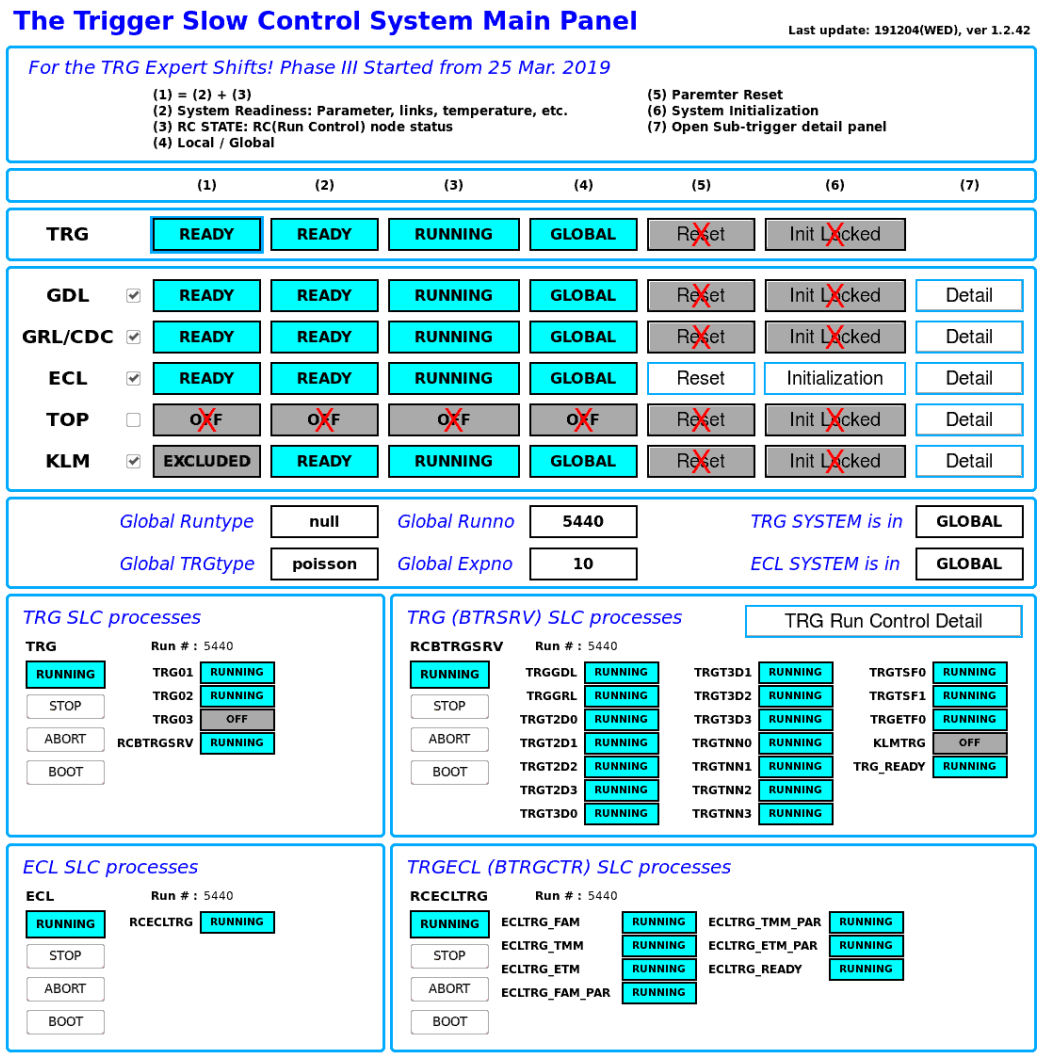}
\caption{The trigger main control panel in operation. If a  run is stable, all run related parts are shown in light-blue color. The KLM trigger and TRG03 are excluded in this example of a test run for maintenance. Depending on purpose of a test run, we can easily exclude/include sub-triggers and COPPER control servers.}
\label{fig:main} 
\end{figure*}

\section{Trigger slow control panels}

\subsection{Main control panel}
\label{sec:TrgMainPanel}

Figure~\ref{fig:main} is a screenshot of the trigger main control panel in a test run during a stable operation. Trigger experts utilize this panel mainly for monitoring the trigger system status. All run related boxes are shown in right-blue color if everything works well during a physics run. When the trigger slow control system detects a problem, it automatically stops the run and shows which parts have the problem on the main panel.

The third block is for integrated trigger information, and the fourth block is for sub-trigger information. Except for the TOP trigger, all sub-triggers are included in the trigger RC scheme (Fig.~\ref{fig:runcon}). The TOP trigger group has a slow control tool that is not based on the Belle II slow control framework, yet. It will also be converted and included in the trigger main RC scheme. The fifth block gives run information: global run type, trigger type, run number, and experimental number. The sixth to ninth blocks correspond to Fig.~\ref{fig:SRV}. Trigger experts can check the status of each node.

We also prepared a read-only version of the panel for the Control Room (CR) shift crews who are the main operators of the Belle II data taking. Opening detail panels and the trigger local run control panel, and reset/initialization buttons are deactivated, but an alarm exists in this version.

\subsection{Level 1 trigger rate plot}
\label{sec:L1}

\begin{center}
\begin{figure*}[!htb]
\centering
\includegraphics[width=0.95\linewidth]{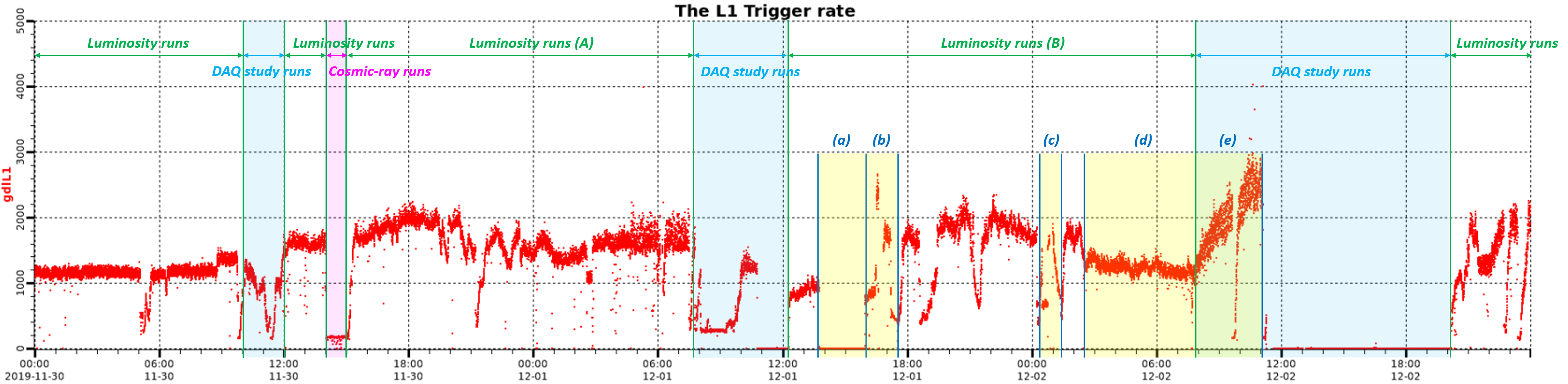}
\caption{The L1 trigger plot. A screen shot of the L1 trigger archived PV displayed on CSS. The x-axis is time, and the range is between November 30, 2019, 00:00 JST and December 3, 2019, 00:00 JST. The unit of the y-axis is hertz ($\mathrm{Hz}$). The L1 trigger rate is between 1 $\mathrm{kHz}$ and 2 $\mathrm{kHz}$ during physics runs.}
\label{fig:L1TRG}
\end{figure*}
\end{center}

\begin{figure*}[!htb]
\centering
\includegraphics[width=0.85\linewidth]{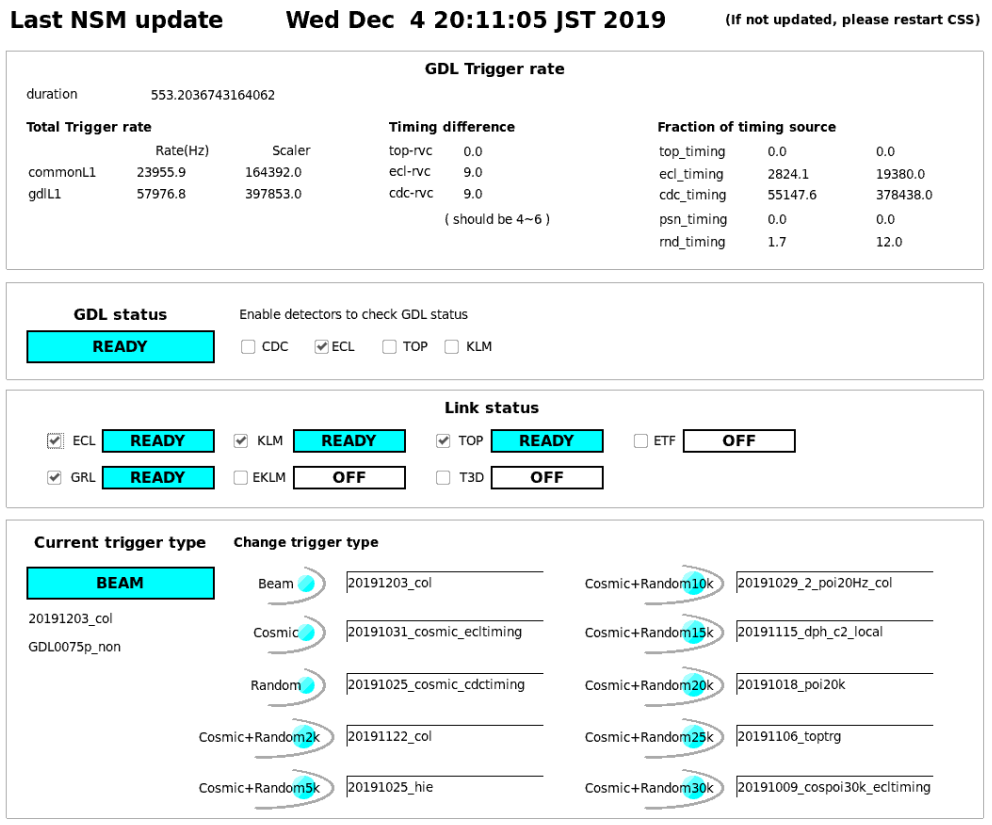}
\captionsetup{justification=centering}
\caption{The GDL detail main panel.}
\label{fig:GDL}
\end{figure*}

The level 1 trigger is the final output of the trigger system, and its rate can be used as one of the direct indicators of the trigger system, so it should be archived and monitored. Figure~\ref{fig:L1TRG} is a screenshot of the L1 trigger rate plot. If the rate suddenly drops or soars without a corresponding change in accelerator conditions, it is suspected that the trigger system has a problem. Then, the CR shift crews should stop a run and call a trigger expert shift crew. The DAQ study runs are mainly for DAQ study, and also for each sub-detector study. During the cosmic-ray runs, the GDL parameter setting is different from the parameters for the physics (luminosity) runs, so the L1 trigger rate is noticeably different from that of physics runs. The luminosity runs in the period (A) in Fig. 6 are relatively stable than that in the period (B) where several problems happened.

\begin{enumerate}[(a)]
    \item No L1 trigger rate, due to beam problem. 
    \item The GDL timing problem happened.
    \item Same as (b)
    \item Sudden decrease of the L1 trigger rate.
    \item ECL expert did local run study in order to chase the origin of (d).
\end{enumerate}

\begin{figure*}[!htb]
\centering
\includegraphics[width=0.95\linewidth]{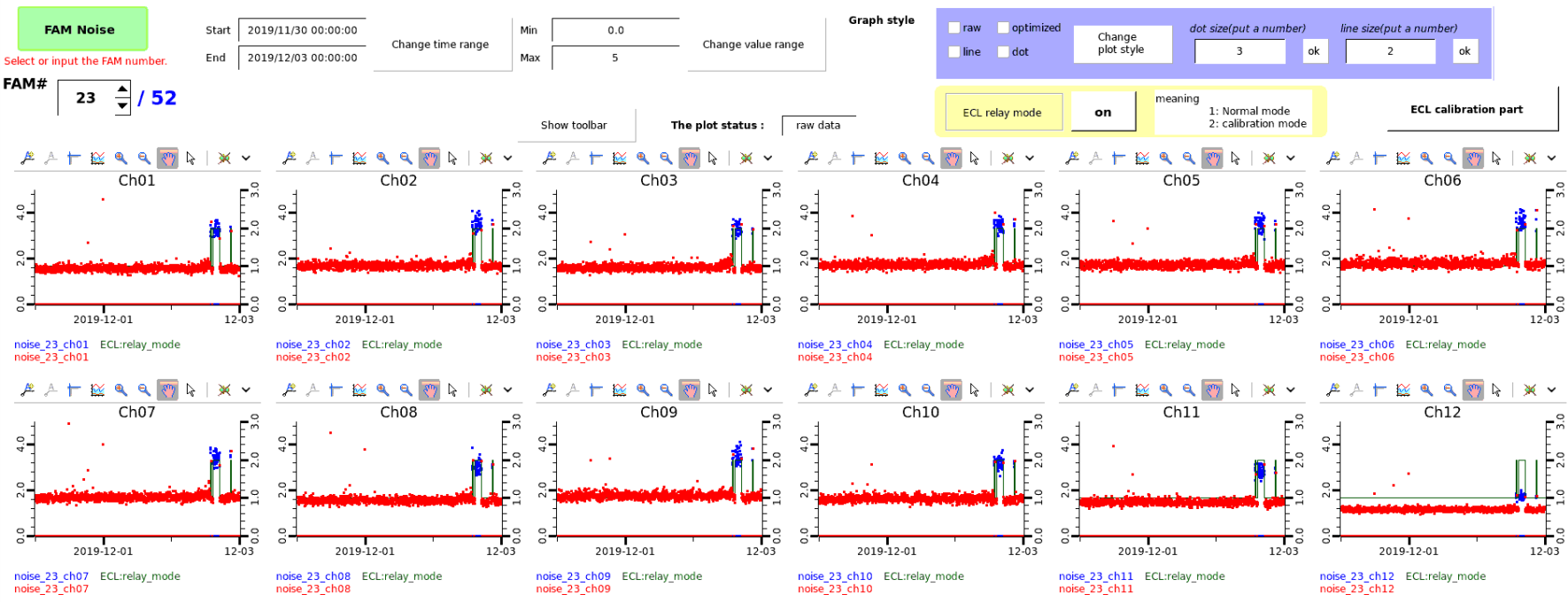}
\caption{The FAM module noise monitoring panel. The x-axis is time: the range is between November 30, 2019, 00:00 JST and December 3, 2019, 00:00 JST, same as Fig.~\ref{fig:L1TRG}. The left-hand side y-axis is for noise value (red/blue dots), and the unit is ADC count (1 ADC count corresponds to 5.25 MeV). The right-hand side y-axis denotes modes of the ECL experimental parameter setting (green line, 1: the normal setting, 2: the local work setting). With the normal setting, the FAM noise level is near 2 ADC counts (red dots), and with the local work setting, the FAM noise level is between 3 and 4 ADC counts depending on the local study (blue dots). The channel 12 is a null-channel. We can change the range of x and y axes, and style of both of the plots by the corresponding setting interface on the panels.}
\label{fig:ECLTRG} 
\end{figure*}

\subsection{Local run control panel}
We regularly take a local run data set for steady improvement of the trigger system, such as calibration or new firmware confirming purposes. For taking a local run data set easily, the trigger local run control panel is prepared. Before a local run, the trigger system should be excluded from the global run. During a local run, status of COPPERs, such as the sizes of the total stored data (bytes) and data-storing rates, are displayed.

\subsection{GDL panel}
Figure~\ref{fig:GDL} is the GDL main panel. Sub-trigger rates and link status of data-taking are displayed on the GDL main panel. With this panel, we can set several GDL parameters for specific purposes such as physics (beam) run, cosmic-ray run, test runs with a random (Poisson) trigger rate, and a cosmic-ray + random trigger rate. We can test the DAQ system stability with an increased random trigger rate.

\subsection{ECL trigger panels}
\label{sec:EcltrgPanels}

The ECL trigger system consists of 3 sub-modules, which are the FADC Analysis Module (FAM), the Trigger Merger Module (TMM), and the ECL Trigger Master (ETM)~\citep{SHK17, BGC10}.

Figure~\ref{fig:ECLTRG} is a FAM noise monitoring panel and the plots in the figure are based on archived PVs. This figure shows an example of how we exploit the trigger slow control system in real use. There are also other panels showing the FAM pedestal, hit-rate, average hit-rate, the TMM temperature, and the ETM physics bits plots. We use all these plots for operating the ECL trigger system stably. Especially, FAM hit-rates and average hit-rates PVs are utilized not only by the trigger group but also by the SuperKEKB side for beam tuning. Additionally, the ECL trigger slow control system provides an automatic parameter setting function depending on run condition~\citep{CH18}.

\section{Trigger slow control in operation}
A read-only version of the trigger main panel and the L1 trigger rate plots are used by CR shift crews. If a run stops automatically by the trigger slow control system or CR shift crews find a problem in the trigger side with the operator interfaces, they contact the trigger expert shift crews by internal chat or a phone call.

If a trigger expert shift crew finds a problem or gets a message/phone call from CR shift crews, the trigger expert first checks of which part shows the problem and follows recovery procedures. If the problem is due to the slow control system itself, the trigger expert can easily recover whole processes on the trigger control servers, COPPER control servers, and all sub-servers of both of them by scripts (Fig.~\ref{fig:SRV}). If the problem is due to a sub-trigger system, the trigger expert can follow the troubleshooting procedure provided by the corresponding sub-trigger group. If the trigger expert cannot recover the systems by following the procedures, then the real expert of the sub-trigger is called.
\section{Summary}

The trigger slow control system has been developed based on the Belle II slow control software framework. The trigger slow control system stores 67,662 configuration parameters into the Belle II central database server for every run start and stop, and 3,303 variables into the Belle II main archiver server every default period of 10 seconds. Shift crews can find problems by intuitive GUI panels and try a recovery procedure when a problematic situation occurs. During the initial running period of Belle II, we have steadily refined the system based on operation experiences. The slow control system made it possible to stably operate the trigger system during the last 2019 autumn run and following runs in 2020.

\section*{Acknowledgement}
This research was supported by Basic Science Research program (NRF-2018R1A2B3003643) and Foreign Large-size Research Facility Application Supporting project (NRF-2019K1A3A7A09033857 and NRF-2019K1A3A7A09035008) through the National Research Foundation of Korea.

\end{document}